\documentclass[twocolumn,showpacs,prb,amsmath,amssymb,citeautoscript,longbibliography]{revtex4-1}
\pdfoutput=1

\usepackage{graphicx}
\usepackage{amsmath}
\usepackage{amsfonts}
\usepackage{amssymb}
\usepackage{color}

\usepackage{hyperref}
\hypersetup{
    bookmarks=false,         % show bookmarks bar?
    unicode=false,          % non-Latin characters in AcrobatÕs bookmarks
    pdftoolbar=false,        % show AcrobatÕs toolbar?
    pdfmenubar=true,        % show AcrobatÕs menu?
    pdffitwindow=false,     % window fit to page when opened
    pdfstartview={FitH},    % fits the width of the page to the window
    pdftitle={Recent progress in many-body localization},    % title
    pdfauthor={Dmitry A. Abanin, Zlatko Papic},     % author
    pdfsubject={},   % subject of the document
    pdfcreator={},   % creator of the document
    pdfproducer={}, % producer of the document
    pdfkeywords={many-body localization} {matrix elements} {disordered systems}, % list of keywords
    pdfnewwindow=true,      % links in new window
    colorlinks=true,       % false: boxed links; true: colored links
    linkcolor=black,          % color of internal links (change box color with linkbordercolor)
    citecolor=blue,        % color of links to bibliography
    filecolor=magenta,      % color of file links
    urlcolor=blue           % color of external links
}

\newcommand{\be}{\begin{equation}}
\newcommand{\ee}{\end{equation}}
\newcommand{\la}{\langle}
\newcommand{\ra}{\rangle}

\begin{document}
\title{Recent progress in many-body localization}
\author{Dmitry A. Abanin}
\affiliation{Department of Theoretical Physics, University of Geneva, Geneva 1211 Switzerland}
\author{Zlatko Papi\'c}
\affiliation{School of Physics and Astronomy, University of Leeds, Leeds LS2 9JT, United Kingdom}
\date{\today}
\begin{abstract}
This article is a brief introduction to the rapidly evolving field of many-body localization. Rather than giving an in-depth review of the subject, our aspiration here is simply to introduce the problem and its general context, outlining a few directions where notable progress has been achieved in recent years. We hope that this will prepare the readers for the more specialized articles appearing in this dedicated Volume of Annalen der Physik, where these developments are discussed in more detail. 
\end{abstract}
%\pacs{}

\maketitle

\section{Introduction}
\label{sec:introduction}

The research into many-body localization (MBL) is driven by the desire to understand the effects of interactions between particles on the stability of the Anderson insulator, a phase characterised by the absence of diffusion in low dimensional disordered systems \cite{Anderson58, FleishmanAnderson,Basko06,Mirlin05}. On a broader level, the study of MBL is part of a more general quest to understand possible outcomes of the quantum evolution of \emph{generic} many-body systems, i.e., those which contain interactions between particles as well as quenched randomness. (Most systems in nature indeed possess both of these features.) In the simplest instance, a many-body system can be assumed to be isolated from any thermal bath, therefore it evolves in time according to the well-known unitary evolution. The problem then reduces to understanding the general outcomes of this evolution and consequently the fate of such systems at long times. 

The motivation to study isolated quantum many-body systems is not purely academic. Experimental advances of the past two decades have led to the realization of synthetic quantum systems, characterized by an unprecedented degree of quantum control and tunability. 
A prominent class of such systems are ultracold atomic gases in optical lattices, which serve as a versatile platform for studying various aspects of many-body physics in a controlled setting (see Ref. \onlinecite{BlochColdAtoms}). Unlike conventional solids, the systems of ultracold atoms are well-isolated from the environment and display a high degree of quantum coherence. This, along with their long intrinsic time scales and a remarkable set of available experimental tools, makes ultracold atomic systems an attractive laboratory for studying non-equilibrium quantum many-body phenomena. These experimental developments continue to stimulate a strong theoretical interest in the quantum dynamics of isolated many-body systems.

In order to describe quantum dynamics, throughout this article we will refer to the \emph{quantum quench} -- a simple, yet general, probe that can be directly implemented in experiments with synthetic systems~\cite{PolkovnikovRMP}. The system, prepared in an initial state $|\psi_0\ra$ at time $t=0$, undergoes unitary evolution with its Hamiltonian $H$. The initial state $|\psi_0\ra$ is typically a relatively simple (for example, non-entangled) state which can be easily prepared in experiment~\cite{Trotzky2012}; from the point of view of generic Hamiltonians, such states typically have high energy density and therefore can be considered to be far from equilibrium. One would like to understand whether unitary evolution results in equilibration and brings the system into an effectively steady state at sufficiently long times. What is the nature of this steady state in different systems? How is it approached? Thus, a fundamental challenge that emerges is to classify many-body systems according to their non-equilibrium properties, such as, e.g., their behavior under a quench. 

A well-known possible outcome of unitary evolution of an isolated quantum system is \emph{thermalization}: at long times every sufficiently small subsystem reaches an effectively thermal Gibbs state\cite{DeutschETH,SrednickiETH,RigolNature}. Thermal states are largely featureless, meaning they can be completely specified by the values of several global conserved quantities, such as the total energy of the system and the total particle number. Thermalization therefore erases the memory of the initial state $|\psi_0\ra$.   Intuitively, this will occur if the system can act as a heat bath for its (sufficiently small) subsystems. Thermalization requires that different parts of the system exchange energy efficiently, such that states with spatially non-uniform energy density can relax to thermal states. Thus, energy transport is necessary and thermalizing systems are expected to be conducting. Often, thermalizing quantum systems are referred to as \emph{ergodic}, because during their evolution they explore all  configurations allowed by the global conservation laws. The properties of states of ergodic systems and the meaning of thermalization in this context are discussed in Section \ref{sec:ETH} below (see also the Review \onlinecite{ANDP:ANDP201600301} in this Volume).

In contrast to thermalizing systems, recent work~\cite{Basko06,Mirlin05} has established MBL as a general mechanism by which quantum systems can avoid thermalization. The localization and the breakdown of ergodicity in MBL systems occur because strong quenched disorder effectively makes energy exchange processes between different degrees of freedom ``off-resonant"~\cite{Basko06,Mirlin05}. As a result, under quantum evolution, an MBL system retains the memory of the local structure of its initial state $|\psi_0\ra$. Due to their lack of ergodicity, MBL phases cannot be described by the conventional statistical mechanics, and this has brought them into the focus of recent theoretical interest. 

As we explain in Sections \ref{sec:anderson} -- \ref{sec:dyn}, significant  progress in describing the properties of MBL phases has been achieved, in particular by applying concepts borrowed from integrable systems and quantum information theory. Most importantly, it has been realized that the ergodicity breakdown in MBL systems is the result of a new kind of integrability: the MBL phase is characterized by a complete set of emergent quasi-local integrals of motion (``LIOMs" or ``l-bits")\cite{Serbyn13-1, Huse13}. The existence of LIOMs leads to a simple and universal description of MBL phases (see Section \ref{sec:liom}), which accounts for their lack of thermalization and several distinct dynamical properties. A more detailed overview of the LIOM picture of MBL phases can be found in articles~\cite{Rademaker_annal,Ros_annal} in this Volume.

Somewhat surprisingly, despite the absence of energy transport, quantum entanglement \emph{does} spread in the MBL phase~\cite{Znidaric08,Moore12}, and even the remote degrees of freedom become entangled under quantum evolution. In a quantum quench setup described above, this leads to the equilibration of a system to a highly non-thermal state. The entanglement properties of MBL eigenstates as well as  the spreading of entanglement can both be explained in the LIOM picture, as we discuss in Sections \ref{sec:ent} and \ref{sec:dyn}. The connections between dynamical properties and transport is reviewed in article \cite{ANDP:ANDP201600362} of this Volume, while articles \cite{ANDP:ANDP201600332, ANDP:ANDP201600318} investigate the dynamical signatures of MBL phases from the recent point of view of ``out-of-time order correlators".

The stability of MBL phases and the existence of a complete set of LIOMs at strong disorder has been theoretically established in a few simple one-dimensional models~\cite{Chandran14, Rademaker16, SBRG, ScardicchioLIOM, Brien16} (in one of them, even at the level of mathematical rigour \cite{Imbrie16}). More recent theoretical efforts have focused on understanding the breakdown of the LIOM picture, as we discuss in Sections \ref{sec:transition} and \ref{sec:symmetry}. A natural setting to explore the breakdown of LIOMs is the transition between the MBL and ergodic phase. While the complete theory of this transition is currently lacking, important steps towards such a theory have been made by the recent real-space renormalisation group studies~\cite{Vosk15,Potter15}. Of particular importance is the understanding of the so-called rare region effects, which determine the physical properties of the system on either side of the transition. The results of these studies are reviewed in articles~\cite{ANDP:ANDP201600384,Vasseur_annal,Agarwal_annal} of the present Volume.  

Alternatively, the character and existence of LIOMs can be strongly affected by the presence of symmetries in the model. It has been understood that symmetries place stringent constraints on whether MBL can occur~\cite{Vasseur15}. On the other hand, if the symmetry admits an MBL phase, the two  can combine to stabilize certain kinds of topological order at finite energy density~\cite{Huse_order,Bauer13}. More generally, quantum information, usually encoded in the low-energy properties of a system, could then remain ``protected" at much higher temperatures, motivating the use of MBL to enhance robustness of quantum information processing schemes \cite{Bahri}.

In Section \ref{sec:exp} we address the experimental status of MBL. Recently, several experimental systems have provided complementary insights into MBL. For example, signatures of MBL have been observed in systems of ultracold atomic gases (in both one~\cite{Bloch15} and two~\cite{Bloch16} spatial dimensions) and trapped ions~\cite{Monroe16}. Furthermore, experiments studying the dynamics of thermalization of spins on NV-centers in diamonds have appeared~\cite{kucsko}. Our review of MBL experiments will be brief and we refer  the interested readers to the original articles.

Finally, in Section \ref{sec:outlook}, we conclude by discussing several open questions and mention some directions that are not  covered in depth by this article, such as the stability of MBL in the presence of dissipation (see the article \cite{ANDP:ANDP201600181} in this Volume), novel types of non-ergodic states (including possible MBL-like states in systems without quenched disorder), and new phases of matter in Floquet systems \cite{Khemani16,Else16,Curt16,Choi16DTC,Zhang2017}.

\section{Eigenstate thermalization hypothesis and the structure of ergodic eigenstates}
\label{sec:ETH}

% from dynamics to highly excited eigenstates; thermal -- ETH; MBL -- ETH breakdown 
Non-equilibrium properties of a many-body system are intimately related to the structure of its highly excited eigenstates. In principle, if the properties of the eigenstates and their corresponding energies are known, time evolution of an arbitrary initial state can be described. For example, in the quantum quench experiment, the initial state $|\psi_0\rangle$ can be expanded in the system's eigenstates, and the time evolution is determined by the coefficients in this expansion, weighted by the phase factors corresponding to the system's eigenenergies. However, this in itself does not explain in a transparent way why thermalization happens. Indeed, since the time evolution is unitary, any information present in the system's initial state remains preserved as time goes on. This naive point of view appears at conflict with one's intuition and experimental findings that many quantum systems indeed reach thermal equilibrium after sufficiently long time and appear to lose memory of their initial configuration.   

Thermalization in ergodic systems is explained by a powerful conjecture regarding the nature of eigenstates -- the eigenstate thermalization hypothesis (ETH)~\cite{DeutschETH,SrednickiETH,RigolNature}. The ETH states that in ergodic systems, the individual excited eigenstates have thermal expectation values of physical observables, which are identical to those obtained using the microcanonical and Gibbs ensembles. The expectation value of a physical observable associated with an operator $\hat O$ is given by the diagonal matrix element $O_{\alpha\alpha}=\la\alpha| \hat O|\alpha\ra$, where $|\alpha\ra$ is an eigenstate of $H$, $H|\alpha\ra=E_{\alpha}|\alpha\ra$. 
Further, to describe how the system approaches the thermal state, Srednicki introduced an ansatz for the matrix elements of physical operators in the basis of system's eigenstates~\cite{Srednicki96,Srednicki99}:
\be\label{eq:ETH}
O_{\alpha\beta}={\mathcal O} (E)\delta_{\alpha\beta} +e^{-S(E)/2} f(E,\omega) R_{\alpha\beta}. 
\ee
The first term describes the diagonal part of the operator in the eigenstate basis, and ${\mathcal O} (E)$ is a smooth function of the energy. The second term describes off-diagonal matrix elements, and $S(E)$ is the thermodynamic entropy at the average energy $E=(E_{\alpha}+E_\beta)/2$, $f(E,\omega)$ is a smooth function of $E$ and the energy difference $\omega=E_{\alpha}-E_\beta$. Finally, $R_{\alpha\beta}$ is a random number, which has zero mean and unit variance. As we discuss below, the function $f(E,\omega)$ determines the relaxation of the physical observable $\hat O$, and is different for different systems and observables. We note  that the ETH ansatz~(\ref{eq:ETH}) for the matrix elements has been verified in several low-dimensional models~\cite{Rigol-FDT,Prelov13,Haque15,Polkovnikov-rev}.

The ansatz (\ref{eq:ETH}) reflects the fact that, in some ways, highly excited states of ergodic systems can be viewed as random vectors in the Hilbert space. Moreover, it predicts the behavior of temporal fluctuations of local observables $O_t \equiv \langle \psi(t) | \hat O | \psi(t)\rangle$. By denoting the infinite time average as 
\be
\bar O = \lim_{t\to \infty} \frac{1}{t}\int_0^t d\tau \; O_{\tau},
\ee
it directly follows that $\overline{(O_t - \bar O)^2} = \mathcal{O}(e^{-S})$. Thus, independent of the initial state $|\psi_0\rangle$, $O_t$ eventually approaches its equilibrium value $\bar O$, and then remains near that value most of the time. In order to find out how the system approaches the equilibrium value, one can calculate the conditional probability to measure $O_t$ given the initial value $O_0$. This probability \cite{Srednicki99} is the Gaussian function of $O_t - C(t) O_0$, with a small variance $\mathcal{O}(e^{-S})$. Here $C(t)$ is determined by the function $f$ in Eq. (\ref{eq:ETH}) and in particular does not depend on  the quantum state of the system. Thus, the behavior of $O_t$ is nearly deterministic and its approach to equilibrium is is controlled by the frequency dependence of the smooth function $f$ in Eq.(\ref{eq:ETH}).

Apart from explaining thermalization in isolated systems, the ETH provides information about quantum entanglement in the eigenstates of ergodic systems. Let us consider an eigenstate $|\alpha\ra$ with energy $E_\alpha$. We partition the system into two subsystems, $A$ and $B$, and ask how strongly $A$ is entangled with $B$ in the state $|\alpha\ra$. Provided $A$ is sufficiently small, such that $B$ can act as an efficient thermal bath, all observables acting on the degrees of freedom in $A$ have thermal expectation values, with the effective temperature $T$ determined by the energy of the eigenstate $|\alpha\ra$: $E_\alpha=\la H\ra_T$. 
This implies that the reduced density matrix of $A$,
\be\label{eq:reduced}
\rho_{A}^\alpha={\rm tr}_B |\alpha\ra \la\alpha| , 
\ee
is equal to the thermal density matrix at temperature $T$:
\be\label{eq:rho_thermal}
\rho_{A}^\alpha=\rho_A (T)=\frac{1}{Z}e^{-\frac{H}{k_B T}}. 
\ee
The amount of entanglement between $A$ and $B$ in the state $|\alpha\rangle$ can be quantified, e.g., using the entanglement entropy, which is the von Neumann entropy of $\rho_A^\alpha$:
\be\label{eq:Sent}
S_{\rm ent}^\alpha(A)=-{\rm tr}_A \left( \rho_A^\alpha \ln \rho_A^\alpha \right).
\ee
The relation (\ref{eq:rho_thermal}) then implies that the entanglement entropy, which depends on microscopic details of a quantum state, is equal to the thermodynamic entropy of $A$, which is determined by the counting of states at temperature $T$. Generally, the latter scales proportionally to the volume of region $A$, and therefore we conclude that the entanglement entropy of a system obeying the ETH also scales with the volume. This reflects the fact that ergodic eigenstates are highly entangled, and agrees with the intuition that the ETH eigenstates are similar to random vectors. As we will see below, non-thermalizing systems such as MBL systems have much lower entanglement in their eigenstates.

\section{From single-particle to many-body localization}\label{sec:anderson}

% brief history
The study of localization began when Anderson introduced the problem of a single quantum particle moving in a disordered crystal in $d$ spatial dimensions~\cite{Anderson58}. A basic model which exhibits Anderson localization is the tight-binding model with the random chemical potential on each site:
\be\label{eq:Anderson}
H_0=J\sum_{\la ij\ra} c_{i}^\dagger c_j +\sum_i \mu_i c_i^\dagger c_i,
\ee
where $c^\dagger, c$ are the creation/annihilation operators, $J$ is the amplitude to hop between nearest neighbor sites $i$ and $j$, and $\mu_i$ are independent random numbers with zero mean and variance $W^2$, such that the typical value of $|\mu_i|$ is of the order $W$. 

The essence of Anderson localization is that at sufficiently strong disorder, the character of the eigenstates changes: instead of extended Bloch waves in a clean crystal ($W=0$), wave functions become exponentially localized around some site ${\bf R}_\alpha$ in the lattice: $|\psi_\alpha ({\bf r})|^2 \propto e^{-|\vec{r}-\vec{R}_\alpha|/\xi_0}$, where $\xi_0$ is the localization length. Such a change of the wave functions leads to the disappearance of diffusion, and the system becomes the Anderson insulator. 

Intuitively,  in the limit of very strong disorder, $W\gg J$, Anderson localization occurs because hopping processes between nearby sites are typically off-resonant, which prevents hybridization of the wave functions on neighboring sites. Further, hopping processes between more remote sites, which arise at higher orders in perturbation theory in $J/W$, are also off-resonant (proving this is a difficult task), and a particle remains localized in some region of space. It turns out, however, that in low-dimensional systems, $d=1,2$, all states are localized even when disorder is weak, $W\ll J$. In $d=3$, all states of the model (\ref{eq:Anderson}) are localized when disorder is sufficiently strong, $(W/J)\geq w_c$. At weaker disorder, there is a mobility edge: states near the band edges are localized, while states around the middle of the band are extended. 

Many theory works over the few decades following Anderson's paper focused on understanding various aspects of single-particle localization. One fundamental challenge that stood out was to understand the effects of interactions on a system in which the single-particle states are localized. In particular, does localization survive when a generic two-body interaction is introduced, and a finite density of particles is considered? This has been a long-standing open question in the field, which was posed already in the original Anderson's paper in 1958~\cite{Anderson58} and later considered by Fleishman and Anderson in 1980 \cite{FleishmanAnderson}. Recent theory works~\cite{Basko06,Mirlin05} have established a positive answer to this question: localization is indeed stable with respect to short-range and sufficiently weak interactions. Such a non-thermalizing phase is called the ``many-body localized" (MBL) phase. In contrast to the Anderson insulator, which is realized in the special case when particles do not interact with each other, the MBL phase can indeed be viewed as a \emph{phase} of matter, i.e., it is robust under sufficiently weak but generic perturbations. The following Sections will highlight some further differences between the MBL phase and the Anderson insulator. 

\begin{figure}[t]
\begin{center}
\includegraphics[width=\columnwidth]{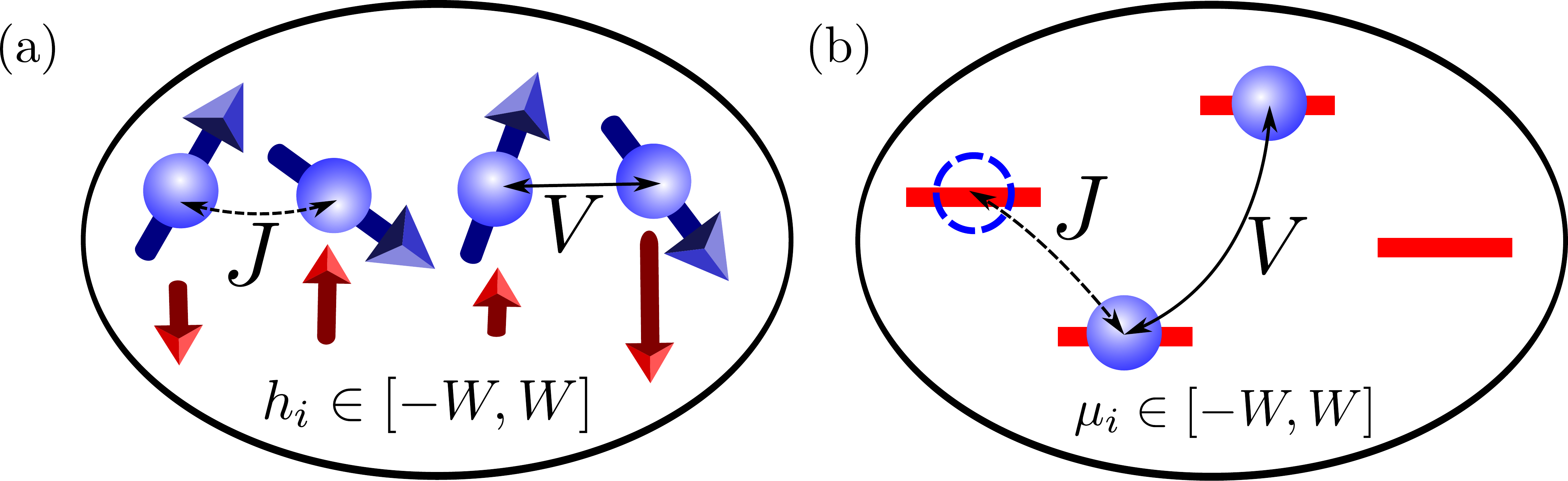}
\caption{ \label{fig:model} (a) A popular model for MBL: a closed, one-dimensional spin $\frac{1}{2}$ XXZ chain, with nearest neighbor hopping ($J$) and interactions ($V$), and a uniform random field $h_i$ pointing along the $z$-axis. (b) In one dimension, the XXZ model (a) is equivalent to a model of spinless fermions on a lattice interacting via nearest neighbor density-density interactions, in the presence of a random on-site chemical potential $\mu_i$. 
}
\end{center}
\end{figure}

In addition to the arguments based on perturbation theory, the existence of the MBL phase has been further supported by extensive numerical simulations of one-dimensional spin and fermionic models~\cite{OganesyanHuse,PalHuse}. Most of these studies have focused on a one-dimensional spin-$1/2$ XXZ model with a random field along the $z$ direction: 
\be\label{eq:Heisenberg}
H= J\sum_{i=1}^{N-1} \left( S_i^x S_{i+1}^x+S_i^y S_{i+1}^y \right) +V\sum_{i=1}^{N-1} S_i^z S_{i+1}^z + \sum_{i=1}^N h_i S_i^z.
\ee
Here $S_i^\alpha = \sigma_i^\alpha/2$ is the Pauli operator acting on site $i$ ($\hbar=1$), $J$ is the hopping amplitude, and $V$ is the interaction strength (see Fig. \ref{fig:model}(a)). The magnitude of the random field is usually chosen to be uniformly distributed $h_i\in [-W,W]$. We have assumed an open chain with $N$ spins. 

The above spin system is equivalent to a model of interacting spinless fermions in a disordered 1d crystal (the mapping between two models is obtained using the Jordan-Wigner transform):
\be\label{eq:fermionic}
H=J\sum_{i=1}^{N-1} c_{i}^\dagger c_{i+1} + h.c. + V \sum_{i=1}^{N-1} \rho_i \rho_{i+1} + \sum_{i=1}^N \mu_i \rho_i,
\ee
where we introduced the density operator $\rho_i = c_i^\dagger c_i-\frac{1}{2}$. The random field $h_i$ translates into the random chemical potential $\mu_i$ for fermions, and the conserved total magnetization in the spin model corresponds to a fixed fermionic filling (e.g., total $S_z=0$ corresponds to half filling).
Compared to the Anderson model (\ref{eq:Anderson}), here the nearest-neighbor density-density interaction is introduced. %(When $V=0$, the model reduces to the non-interacting Anderson insulator.) 
We note that several other models have also been considered in the literature, e.g., models with bond disorder \cite{Vasseur15}, disordered quantum Ising model \cite{Kjall14}, etc. 

One important feature of finite lattice models like (\ref{eq:Heisenberg}) or (\ref{eq:fermionic}) is that their spectra are bounded. (For a discussion of MBL in systems with unbounded spectra, see the article \cite{ANDP:ANDP201600365} in this Volume.) This facilitated the microscopic investigations of MBL phases, in particular via numerical diagonalization of small 1d systems. 
Numerical works have established that all states of the model (\ref{eq:Heisenberg}) are indeed many-body localized when disorder is sufficiently strong, $W>W_c$. For $J=V$ (Heisenberg model), critical disorder strength was found to be $W_c\approx 3.5$ \cite{PalHuse}. In most theory work preceding these numerical investigations, the focus was on direct evaluation of the conductivity in the MBL phase, with the goal of demonstrating the absence of transport. Numerical simulations of lattice models have instead probed the existence of the MBL phase by  examining the properties of individual many-body eigenstates. This approach was particularly useful in understanding the universal properties of MBL phases, as we review in the following Section.

\section{Many-body localization: local integrals of motion}\label{sec:liom}

A powerful insight into the physics of the MBL phase, brought about by the microscopic analysis of its eigenstates, has been the theory of ``local integrals of motion" (LIOMs) \cite{Serbyn13-1, Huse13}. We now outline the main steps leading to this phenomenological description of MBL states. We assume strong disorder such that all of the MBL eigenstates are localized. For concreteness, let us consider model (\ref{eq:Heisenberg}) with $N\gg 1$ spins, although the arguments are applicable to a variety of models. In the classical limit $J=0$, the system is trivially localized because the eigenstates are simply product states, with each spin pointing either up or down, i.e., $|\alpha_0\ra=|\{ \sigma_i^z \}\ra=|\sigma_1^z \sigma_2^z...\sigma_N^z \ra$, $\sigma_i^z=\uparrow,\downarrow$. The Hamiltonian commutes with the $z$-projection of spin at every site: $[\sigma_i^z, H]=0$, and therefore there exists a complete set of mutually commuting, strictly local integrals of motion $\sigma_i^z$ with eigenvalues $\pm 1$. 

Next, let us turn on the coupling $J>0$, which leads to spin-flip processes, while keeping disorder sufficiently strong, $W>W_c(J)$, such that all states remain MBL. Intuitively, it is expected that MBL eigenstates $|\alpha\ra$ at $J>0$ remain, in some sense, close to the  product states $|\alpha_0\ra$ at $J=0$. More precisely, an eigenstate $|\alpha\ra$ is related to $|\alpha_0\ra$ by a quasi-local unitary transformation $U$ which creates spin flips only between nearby degrees of freedom (the long-distance flips are exponentially suppressed). 
The quasi-local unitary transformation $U$ diagonalizes the Hamiltonian: 
\be\label{eq:U}
U^\dagger H U=H_{\rm diag}, 
\ee
where $H_{\rm diag}$ is diagonal in the up-down basis $|\{ \sigma_i^z \}\ra$. The transformation $U$ can be constructed perturbatively in the small parameter $\lambda=J/W\ll 1$. 

\begin{figure}[t]
\begin{center}
\includegraphics[width=\columnwidth]{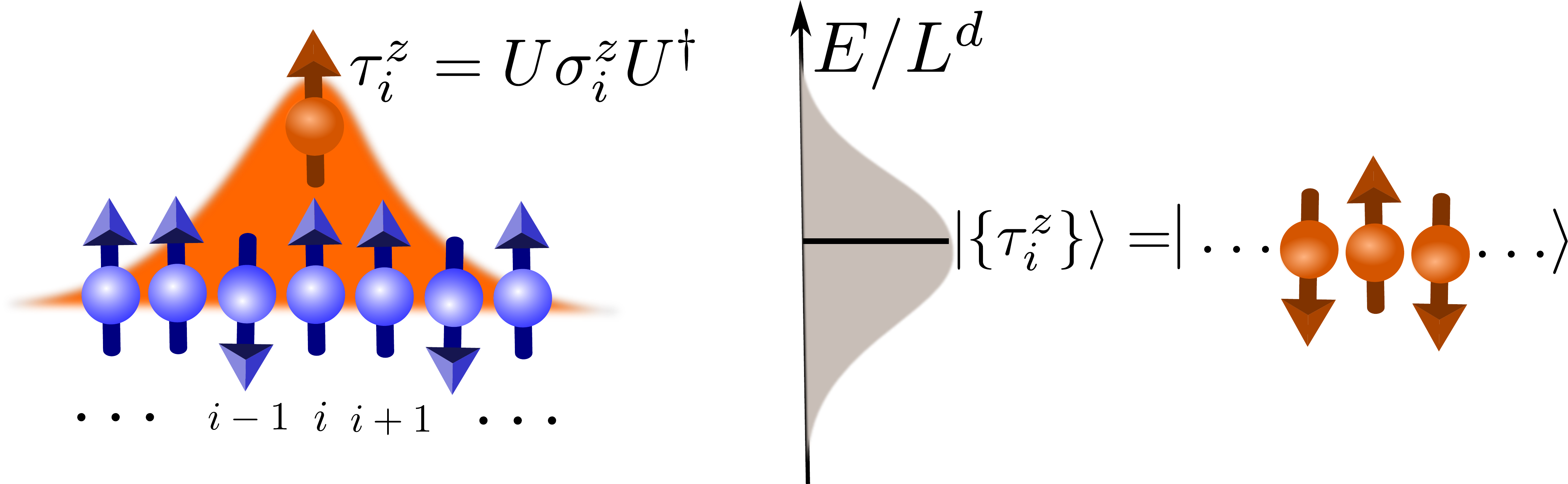}
\caption{ \label{fig:liom} In the MBL phase at sufficiently strong disorder, there emerges an extensive number of LIOMs, $\tau_i^z$. Each $\tau_i^z$ is a Pauli operator unitarily related to the original spins, $\tau_i^z = U \sigma_i^z U^\dagger$, with support decaying exponentially away from the site $i$. Different $\tau_i^z$ commute with each other, as well as with the Hamiltonian $H$. Each eigenstate of the system is completely specified by the simultaneous quantum numbers of all $\{ \tau_i^z \}$.
}
\end{center}
\end{figure}

Observing that $H_{\rm diag}$ commutes with every $\sigma_i^z$ operator, we can now introduce a set of integrals of motion for $H$:
\be\label{eq:tau}
\tau_i^z=U\sigma_i^z U^\dagger. 
\ee
It is easy to see from (\ref{eq:U}) that $\tau_i^z$ operators commute with the Hamiltonian and with each other:
\be\label{eq:commute}
[\tau_i^z,H]=[\tau_i^z,\tau_j^z]=0. 
\ee
Formally, such operators can be defined for any quantum system. However, the crucial property of the MBL phase is the quasi-locality of the transformation $U$, which makes $\tau_i^z$ quasi-local operators -- that is, their support is mostly concentrated around one site and decays exponentially away from it, see Fig. \ref{fig:liom}. For example, in the model of Eq. (\ref{eq:Heisenberg}), the explicit form of $\tau_i^z$ is 
\be\label{eq:tauexpansion}
\tau_i^z \approx \sigma_i^z + \sum_{j,k} \sum_{a,b=x,y,z} f_{i;jk}^{ab} \sigma_j^a \sigma_k^b + \ldots,
\ee
where the weights decay exponentially with distance,
\be\label{eq:f}
f_{i;jk}^{ab} \propto \exp(-\max\{ |i-j|,|i-k|\}/\xi),
\ee
and the dots $\ldots$ denote higher order ($N\geq 3$) spin terms. 
Observe that certain terms in the expansion (\ref{eq:tauexpansion}) may vanish for symmetry reasons (e.g., there is no $\sigma_i^y$ term, etc.).

The spatial decay of $\tau_i^z$ away from site $i$ can be used to define a ``many-body localization length", which we denoted by $\xi$ above. Unlike the Anderson case, in MBL systems there could be several characteristic length scales that determine various properties of the MBL phase (moreover, these length scales will fluctuate depending on the eigenstate). For example, in Section \ref{sec:dyn} we will encounter another length scale, $\tilde \xi$, which is in principle different from $\xi$, and controls the dephasing dynamics in the MBL phase. These various length scales may also behave differently as the MBL phase is driven towards the transition to the thermal phase. In particular, some of the lengthscales, like $\xi$, might be expected to diverge at the transition point, while others (e.g., $\tilde \xi$) could remain finite (see Ref. \onlinecite{Huse13} for more details). Finally, in the thermal phase, $U$ is highly non-local, there is no exponential hierarchy in the $f$ coefficients in Eq. (\ref{eq:tauexpansion}), and consequently $\tau_i^z$ are not very useful. 

%LIOMs--> erg breaking
The operators $\tau_i^z$ are usually referred to as LIOMs~\cite{Serbyn13-1} or l-bits~\cite{Huse13}. They are Pauli operators with eigenvalues $\pm 1$, and form a complete set: specifying the values of $\tau_i^z=\pm 1$ for all $i$ uniquely specifies an eigenstate $|\alpha\ra=|\{\tau_i^z\}\ra$. The emergence of such quasi-local conservation laws provides an intuitive explanation for the ergodicity breaking in the MBL phase: indeed, under unitary evolution the expectation value of each $\tau_i^z$ is conserved, and therefore the system retains the local memory of the initial state at arbitrarily long times. 

The relation (\ref{eq:tau}) defines the operators $\tau_i^z$ in terms of the physical spin operators $\sigma_i^\alpha$, $\alpha=x,y,z$. In order to express an arbitrary physical operator in terms of $\tau$-spins, we introduce the operators $\tau_i^{x(y)}$ as $\sigma_i^{x(y)}$ dressed by the transformation $U$:
\be\label{eq:taux}
\tau_i^{x(y)}=U\sigma_i^{x(y)} U^\dagger. 
\ee
Operators $\tau_i^\alpha$, $\alpha=x,y,z$ and their products form a basis in the operator space, and any physical operator can be expanded in this basis. It is worth noting that the relation between physical operators and $\tau_i^\alpha$ operators is quasi-local. By analogy, one can further define $\tau_i^{\pm}$, the raising/lowering operators for the effective spin $i$.

The above picture of the MBL eigenstates has been supported by various  studies which constructed LIOMs explicitly~\cite{Chandran14, Rademaker16, MonthusLIOM, SBRG, Pollet16, Brien16}. Ref.~\onlinecite{ScardicchioLIOM} established the existence of LIOMs using the perturbative techniques of Ref.~\onlinecite{Basko06}. Finally, Ref.~\onlinecite{Imbrie16} provides a mathematical proof of the quasi-locality of the unitary $U$ in a certain 1d MBL system, under some natural assumptions regarding the spectral properties (the absence of level attraction). As we discuss in the following Sections, the theory based on LIOMs allows one to understand the 
entanglement properties and dynamics in the MBL phase, in particular the spreading of entanglement in the quantum quench setup, which was first observed in numerical simulations~\cite{Znidaric08,Moore12}. 

The local integrals of motion in the MBL phase are discussed in depth in reviews~\cite{Rademaker_annal,Ros_annal} in this Volume.

\section{Entanglement and classical simulations of many-body localized states}\label{sec:ent}

At strong disorder, all eigenstates in the MBL phase are product states of LIOMs $\{ \tau_i^z\}$. Each LIOM $\tau_i^z$ is related to the physical spin at the site $i$, up to spin-flip corrections which are exponentially suppressed in the distance from $i$, see Eq. (\ref{eq:tauexpansion}). In this Section, we explain that 
such a form of the eigenstates leads to strong constraints on their entanglement properties, which in turn has implications for their simulations on classical computers.

Assuming the existence of LIOMs, it is intuitively clear that in a 1d MBL system, the entanglement entropy $S_{\rm ent}$ of the eigenstates is bounded by a constant. Consider a finite chain of length $L$ bipartitioned in the middle. The only contributions to the entanglement  entropy come from the terms in the expansion of $\tau_i^z$, see Fig. \ref{fig:liom} and Eq. (\ref{eq:tauexpansion}), located near the bipartition, i.e., within the length $\xi$ introduced in Eq.(\ref{eq:f}). As we increase $L$ to the thermodynamic limit, we expect the entropy to saturate to a constant $S_{\rm ent}(A)\approx C$ when $L\gtrsim \xi$. This was verified in numerical simulations of several 1d models~\cite{Bauer13, Serbyn13-1, Kjall14}. In higher dimensions, the entropy of the subsystem $A$ is proportional to the number of degrees of freedom at the boundary of the region, $\partial A$. Thus, MBL eigenstates generally obey a very low, boundary-law scaling of entanglement entropy \cite{Eisert08}. The boundary-law should be contrasted with that of the (excited) eigenstates of ergodic systems, which exhibit a much larger, volume-law entanglement, as we saw in Section \ref{sec:ETH}. 

The boundary-law entanglement of (excited) MBL eigenstates makes them similar to the ground states of 1d gapped systems. In the latter case, 
the spectral gap is responsible for the exponential decay of correlations \cite{Hastings2006, Nachtergaele2006, Hastings07}. By contrast, in MBL systems the boundary-law for entropy is a consequence of the existence of LIOMs. This guarantees the boundary law scaling in arbitrarily high excited eigenstates, despite the fact that the splitting between the many-body energy levels vanishes exponentially ($\sim WL/2^{L}$ in the spin-1/2 example). 

Due to the boundary law, MBL states should be 
amenable to efficient classical numerical simulation. Indeed, it is known that states with such low entanglement can be represented in a compact way, using the number of parameters which scales polynomially with the number of degrees of freedom~(for a review, see, e.g., Ref.~\onlinecite{Eisert08}). This implies that MBL eigenstates can be efficiently simulated classically using tensor network formalism~\cite{Verstraete_review}. This observation has opened the door to studying MBL numerically in large systems. 

Recent works~\cite{Pekker15,LimSheng, KhemaniPollmann16, Karrasch15, Serbyn_spectrum} developed extensions of the density matrix-renormalization group (DMRG) algorithm \cite{White}, typically used for ground states of 1d quantum systems, which allow one to obtain individual, highly excited MBL eigenstates. Further, it was argued that the existence of a complete set of LIOMs implies an efficient representation of all eigenstates with a single spectral tensor network~\cite{Chandran_spectral}. Variational algorithms for diagonalizing MBL Hamiltonians using tensor networks have been recently developed~\cite{PollmannCirac16,Pal17}. 

In addition to the entanglement entropy, some recent works have explored the structure of the \emph{entanglement spectrum}~\cite{Haldane08} in MBL \cite{Monthus, Rahul16, Serbyn_spectrum} and ergodic systems \cite{Chamon15}. The entanglement spectrum, i.e., the eigenvalue spectrum of the reduced density matrix, reveals further information about the system which may not be directly accessible in the entanglement entropy. While the entanglement spectrum of ergodic states was found to obey the Marchenko-Pastur distribution \cite{MarcenkoPastur, Chamon15} (in agreement with random matrix theory), in MBL states the entanglement spectrum obeys a power-law distribution with an exponent determined by the localization length $\xi$ \cite{Serbyn_spectrum}. Thus, the entanglement spectrum reveals a difference between MBL states and ground states of various integrable (or even non-integrable) models \cite{Okunishi}, where the entanglement spectrum decays faster than a power law, even though the entropy in both cases may obey the boundary-law scaling.

\section{Dynamics in many-body localized phases}\label{sec:dyn}

The existence of LIOMs underpins the non-trivial dynamical properties of MBL phases. Here we review these properties, focusing mostly on the quantum quench setup. To understand the mechanism of the dynamics, we first note that the Hamiltonian takes a very simple form in terms of $\tau$-operators: since $H$ commutes with every $\tau_i^z$, it can only involve $\tau_i^z$ operators and their products~\cite{Huse13,Serbyn13-1,Serbyn13-2}, 
\be\label{eq:Htau}
H=\sum_i \tilde h_i \tau_i^z+\sum_{ij} J_{ij} \tau_i^z \tau_j^z + \sum_{ijk} J_{ijk} \tau_i^z \tau_j^z \tau_k^z+\dots
\ee
The couplings between remote LIOMs decay exponentially with the distance between them, e.g., 
\be\label{eq:decay}
J_{ij}\propto J_0 \exp(-|i-j|/{\tilde \xi}), 
\ee
where ${\tilde \xi}$ is a characteristic length scale that controls the dephasing dynamics, as we will see below. The exponential decay of the interactions stems from the fact that in the physical basis, $H$ is a sum of local terms, and $\sigma_i^\alpha$ operators are quasi-local in terms of $\tau$-operators. 

%dynamics - dephasing, exponentially decaying interactions are important. Entanglement and quantum correlations, diagonal ensemble 
The Hamiltonian (\ref{eq:Htau}) describes simple dynamics of $\tau$-spins: each effective spin is precessing in the magnetic field $h_{i,\rm eff}$ directed along the $z$ axis, created by other spins. The $\tau_i^z$ component of the spin $i$ is conserved during the evolution. Importantly, the magnetic field $h_{i,\rm eff}$ experienced by the spin $i$, depends on the direction of other spins $j\neq i$. Therefore the angle of rotation of the $i$th spin depends on the state of all other spins. 
If the system was prepared in a superposition of sufficiently many eigenstates (which is true for a generic initial state), this would generate entanglement between remote effective spins. Two spins situated at a distance $r$ from each other become entangled over a time which is inversely proportional to the interaction coupling them: 
\be\label{eq:dephasing}
\tau_{\rm ent}(r)\sim \frac{\hbar}{J_0 e^{-r/{\tilde \xi}} }=\frac{\hbar}{J_0} e^{r/{\tilde \xi}}. 
\ee
Therefore, quantum correlations propagate logarithmically in time, $r(t)\propto \ln (J_0t/\hbar)$. This should be contrasted with ergodic systems, where quantum correlations propagate linearly in time~\cite{Chiara06,Kim13}. 

%dephasing physics -- log-growth of entanglement 
In a quantum quench setup, the slow dephasing results in a logarithmic spreading of entanglement: if the system is prepared in an arbitrary non-entangled product state $|\psi_0\ra$, the (disorder-averaged) entanglement entropy of a subsystem $A$ evolves as~\cite{Serbyn13-2,Huse13,Znidaric08,Moore12}:
\be\label{eq:Sent_growth}
S_{\rm ent}(t)\propto {\tilde \xi} \ln(J_0 t/\hbar). 
\ee
This law of entanglement spreading is often viewed as one of the main characteristics of the MBL phase. At long times, $S_{\rm ent}$ saturates to a value $S_{\rm ent}(\infty)$, which is proportional to the volume of the subsystem $A$. Thus, at infinite time, the entanglement of an MBL system would be extensive, but typically smaller by some constant factor from the full thermal entropy of region $A$ (which would be reached in an ergodic system).

Further, the random phases generated during the evolution lead to the {\it equilibration} of all local observables: at $t\to \infty$, $\la \psi(t)| \hat O | \psi(t)\ra\to O_{\infty}$.
It has been shown that the approach of local observables to their equilibrium values occurs in a power-law fashion, with an exponent set by  the parameter ${\tilde \xi}$~\cite{Serbyn14}. This provides an unambiguous signature of the dephasing physics in the MBL phase. Thus, despite the absence of energy transport, MBL systems generally equilibrate, but to a highly non-thermal state, which has the memory of the initial conditions.
It should be noted that the dephasing mechanism described above distinguishes the MBL phase from the non-interacting Anderson insulator, where no equilibration in the quantum quench setup occurs. We note that other ways of observing the logarithmic spreading of correlations have been proposed, including revivals of local observables~\cite{Vasseur14}, and modified spin-echo experiments~\cite{Serbyn_14_Deer}.

\section{Many-body localization transition and Griffiths effects} \label{sec:transition}

Thus far we have discussed the properties of eigenstates and the  dynamics in the MBL phase at strong disorder. How does the transition from the MBL to the ergodic phase occur, when disorder strength is reduced? The MBL transition is an unusual kind of a dynamical phase transition, across which the nature of eigenstates changes dramatically: for example, the entanglement entropy of individual eigenstates changes its scaling from the boundary-law to the volume-law. In the vicinity of the transition, depending on the disorder realization, the system will be a mixture of localized and thermalizing subsystems. Thermalizing regions have the ability to thermalize the localized regions, which is, however, limited due to their finite size.  

The successful scaling theory of localization transitions in single-particle systems~\cite{Scaling_theory} is based on a single parameter, the Thouless conductance~\cite{Thouless72}, which characterizes the response of eigenstates to the change in the boundary conditions. A natural possibility is that the MBL transition, similarly, is a one-parameter scaling theory. Ref.~\onlinecite{Serbyn15} introduced a possible many-body extension of the Thouless conductance, which characterizes how the many-body eigenstates in the vicinity of the MBL transition respond to a local perturbation of the Hamiltonian.  Numerical analysis~\cite{Serbyn15} shows that such a many-body Thouless conductance, defined in terms of matrix elements of local operators, provides a good diagnostic of the MBL transition. In particular, it allows one to map out the position of the many-body mobility edge (the position of the MBL transition depends on the energy density), in agreement with numerical simulations \cite{Alet14}. Related studies of the mobility edge are reported in articles~\cite{ANDP:ANDP201600356,ANDP:ANDP201600399} of the present Volume. An open question is whether the MBL transition is indeed described by a single-parameter scaling, and further investigations are needed.

A microscopic theory of the MBL transition is currently lacking, thus our discussion on this subject will be brief. An important step was taken in Refs.~\cite{Vosk15,Potter15}, which developed a phenomenological real-space renormalization group (RG) description of the MBL transition in one-dimensional systems. These RG procedures involve a set of heuristic rules for merging thermal and MBL subsystems,  which capture the competition between thermalization and localization. Interestingly, on either side of the transition, rare-region (Griffiths) effects become important and determine various physical properties. On the thermal side of the transition, rare MBL regions lead to the sub-diffusive particle number transport, and sub-ballistic entanglement spreading in the quantum quench experiment~\cite{Vosk15,Potter15,Agarwal15}. The anomalous diffusion exponent vanishes continuously at the transition. At the critical point, both particle number and entanglement spread logarithmically in time. The Griffiths effects on transport are expected to be most pronounced in 1d systems, where inclusions of the MBL phase act as bottlenecks for transport. These, and other related recent developments are reviewed in detail in articles~\cite{Vasseur_annal,Agarwal_annal} of the present Volume.

\section{Symmetries and localization-protected quantum order} \label{sec:symmetry}

In the preceding discussion, symmetry has not played a crucial role for the phenomenology or the existence of the MBL phase. (Models like XXZ do have the conservation of energy and total magnetization, but those can be broken without destroying MBL.) We now consider the cases where symmetry has an impact on MBL and thermalization, and discuss the notion of localization-protected quantum order. 

In ergodic systems, the individual eigenstates are effectively thermal, and therefore they exhibit ordering (e.g., break a symmetry of the Hamiltonian) only if ordering exists in the thermodynamic equilibrium. Since MBL eigenstates are not thermal, they can be ordered even if in the thermodynamic equilibrium the system is not ordered. This leads to an exciting possibility of using MBL to protect various kinds of quantum order, including symmetry-breaking and topological order at finite energy density~\cite{Huse_order,Bauer13}. This phenomenon was discussed in Refs.~\onlinecite{Huse_order,Pekker14,Vosk_ising} which considered MBL in random transverse-field Ising spin chains. Such systems are characterized by a global $\mathbb{Z}_2$ symmetry. Depending on the system's parameters, two MBL phases can emerge: a spin-glass-like phase, in which the eigenstates break the $\mathbb{Z}_2$ symmetry, and a paramagnetic phase, in which the eigenstates respect the symmetry. It was argued that the transition between the two distinct MBL phases in this model is described by the infinite randomness fixed point, similar to the ground state phase transition~\cite{Pekker14,Vosk_ising}. Even more interestingly, MBL can protect certain kinds of topological~\cite{Huse_order,Bauer13} and symmetry-protected topological order~\cite{Chandran_SPT} at finite energy density.

The character of MBL eigenstates, and the very existence of the MBL phase, strongly depend on the kind of symmetries present in the system. Either symmetric or symmetry-breaking MBL phases are possible for the case when the symmetry group is Abelian (e.g., $\mathbb{Z}_2$). 
The case when the symmetry group of the Hamiltonian is non-Abelian is qualitatively different. For {\it discrete} non-Abelian symmetries (such as the permutation group $S_n$), MBL is possible, however the eigenstates must spontaneously break the symmetry~\cite{Vasseur15}. Continuous non-Abelian symmetries appear to prohibit MBL altogether, as shown in recent works~\cite{Vasseur16,Protopopov16}. Such systems are always thermalizing, even though thermalization processes can be non-trivial and parametrically slow. Intuitively, non-Abelian symmetries are unfavorable for localization because the eigenstates come in degenerate multiplets; such degeneracies lead to resonances and dictate that boundary-law entanglement scaling is violated. 
%{\color{red} Move to outlook? $\to$} However, an interesting open question concerns the possible existence of non-ergodic phases, which are neither fully MBL nor fully thermal. Such phases might be characterized by scaling of entanglement which is intermediate between boundary-law in MBL systems, and volume-law in thermal systems.

\section{Experimental developments} \label{sec:exp}

On the experimental side, synthetic quantum systems, being isolated from environment, are ideally suited for probing the dynamics of isolated many-body systems, and investigating quantum thermalization and MBL. Recently, signatures of MBL have been observed in one- and two-dimensional systems of ultracold atoms in disordered optical lattices~\cite{Bloch15,Bloch16}, and also in small systems of trapped ions~\cite{Monroe16}. These experiments, in essence, implement a quantum quench setup, where the initial state is a charge-density-wave~\cite{Bloch15}, with occupation numbers alternating between $0$ and $1$ at odd or even lattice sites. The evolution of this state was monitored, and the memory of the initial density modulation at long times was interpreted as indication of MBL. Another promising system for studying quantum many-body dynamics include spins in the NV-centers in diamond, where critical thermalization arising from long-range interactions in this system has been observed~\cite{kucsko}. Further, a recent experiment~\cite{Cappellaro16} reported signatures of MBL and logarithmic spreading of correlations in a system of nuclear spins. 

One of the main challenges for the experiments described above is that synthetic systems are not fully isolated from the environment, and there are slow extrinsic processes which affect the state of the system and typically destroy MBL. Thus, such experiments cannot access intrinsic processes which are extremely slow. Despite this limitation, in the systems of ultracold atoms~\cite{Bloch15,Bloch16} the extrinsic processes are sufficiently slow such  that clear signatures of MBL could be observed. The tunability of these systems can be utilized to study interesting aspects of quantum dynamics, including logarithmic spreading of entanglement and equilibration of local observables in the MBL phase, as well as the effects of symmetries and symmetry-protected order, in a controlled setting. These experiments have already entered a stage where the corresponding model systems cannot be treated numerically using exact diagonalization, and they are expected to bring further interesting discoveries in the near future.

\section{Outlook and some open questions}\label{sec:outlook}

As we discussed above, a rather complete picture of the MBL phase has emerged in recent years, supported by both analytic and numerical results at sufficiently strong disorder. Key properties of MBL phases, which physically distinguish them from Anderson insulators as well as thermal phases, have been identified on the level of individual eigenstates or as a response to dynamical probes. However, despite rapid progress, several fundamental questions remain open. 

An important goal is to develop a more complete theory of the MBL-thermal transition. Currently, the essential roadblock seems to be the lack of a microscopic theory describing the effect of a finite thermalizing subsystem on an adjacent MBL subsystem. The progress on this question will likely feed into better understanding of the stability of MBL phases in higher dimensions, particularly in 2d. Experiments in cold atoms \cite{Bloch16} are already probing this question, but limitations to short time scales make it difficult to draw conclusions about the existence of MBL in 2d. Similarly, in order to interpret experiments on NV centers, a better theoretical understanding of the stability of MBL in the presence of long range (dipolar) interactions \cite{Burin2015, YaoDipolar} is needed. Another set of questions of immediate experimental relevance concerns the stability of MBL phases in the presence of a bath \cite{NandkishoreBath, HuseZeno, JohriBath, NandkishoreProximity, Levi2016}.

%This problem, and more generally the MBL transition, are reviewed in detail in this Volume~\cite{Vasseur_annal,Agarwal_annal}.
%It is intuitively clear that the thermal subsystem will partially thermalize the MBL part, destroying the conservation of nearby LIOMs; if 

Going beyond the conventional MBL is the question on the possible existence of other non-thermalizing phases of matter. Is MBL the only mechanism to break ergodicity? Do eigenstates of non-ergodic systems necessarily have boundary-law entanglement? One possible route to making progress in this direction was suggested recently in Ref.~\onlinecite{Protopopov16}. The basic idea is to consider different entanglement patterns of eigenstates, which emerge for (fine-tuned) ``fixed-point" Hamiltonians, and then study their stability with respect to local perturbations. Such systems, if found, may exhibit a partial, rather than complete (as in MBL) set of quasi-local integrals of motion. In a different direction, Ref.~\onlinecite{Altshuler16} argues the existence of an intermediate, delocalized, but non-ergodic phase, in a (single-particle) hopping problem on a random regular graph (RRG). A different study~\cite{Tikhonov16}, on the other hand, argues that the delocalized states in this model are ergodic. We also note that while the hopping problem on RRG has certain similarities with the problem MBL, it is non-local, and locality appears to be a crucial aspect of MBL. 

Understanding whether quenched disorder is a necessary ingredient for ergodicity breaking is another open problem. Recent works~\cite{Huveneers13,Muller} introduced several translation-invariant models which are expected to show some form of non-ergodicity, or at least slow thermalization. These models are constructed in such a way that most of the transitions/hopping processes which the particles can make, are off-resonant. Thus, the basic physical mechanism is quite similar to MBL. However, it was later argued that the translation invariance may inevitably lead to delocalization and thermalization, mediated by rare resonant ``bubbles" which are mobile~\cite{DeRoeck}. Studying such models numerically has proved to be challenging~\cite{SchiulazMuller15,Yao14,Garrahan15} because of severe finite-size effects, as discussed in Ref.~\onlinecite{Papic}. More recently, a non-generic class of models are being investigated~\cite{MoessnerTI}, following the idea of mapping disorder to an ancillary degree of freedom~\cite{ParedesDisorder, Andraschko2014}. Even if it turns out that in generic models translation invariance always leads to thermalization, it is clear that such models would provide interesting examples of slow thermalizing dynamics. Moreover, similar kinds of models might realize a new kind of matter called ``quantum disentangled liquids"~\cite{QDL, QDL2, QDLEssler}. Such phases might occur in a system of two or more species of particles, where each species thermalizes, thus the system is not MBL, but there remain subthermal interspecies correlations in arbitrarily high energy densities. 

Finally, we briefly mention another set of very recent developments, related to MBL and new phases of matter in the so-called Floquet systems, in which the Hamiltonian is periodically varying in time. Such systems are naturally realized in experiments with synthetic systems; in particular, in recent years periodic driving has emerged as a useful tool for engineering interesting (e.g., topological) band structures for ultracold atoms in optical lattices~\cite{Eckardt_review}. Energy conservation is broken in Floquet systems, and generally they are expected heat up, absorbing energy from the drive and heating up to a featureless infinite-temperature state~\cite{Lazarides14,Alessio14,Ponte14}. If this were always the case, no distinct ``Floquet phases of matter" would exist. However, it has been shown that MBL is possible in disordered Floquet systems, giving rise to a Floquet-MBL phase~\cite{Ponte15,Lazarides15,Abanin20161}. In such a phase, the system fails to thermalize indefinitely: instead, it exhibits phenomenology similar to static MBL systems, including a complete set of LIOMs and slow spreading of entanglement. In a set of recent exciting developments, it was demonstrated that Floquet-MBL systems come in many flavors~\cite{Khemani16}; in particular, a ``Floquet time crystal" phase has been introduced~\cite{Khemani16,Else16,Curt16}. The signatures of this phase have been observed in recent experiments with NV-centers in diamond~\cite{Choi16DTC} and trapped ions~\cite{Zhang2017}. These developments have been recently reviewed in more detail in Ref.~\onlinecite{MoessnerSondhi}.

\section{Acknowledgements}

We would like to thank our collaborators, Thomas O'Brien, Anushya Chandran, Soonwon Choi, Eugene Demler, Sarang Gopalakrishnan, Wen Wei Ho, Fran\c cois Huveneers, Isaac Kim, Michael Knap, Chris Laumann, Misha Lukin, Alexios Michailidis, Pedro Ponte, Tomaz Prosen, Ivan Protopopov, Wojciech de Roeck, Miles Stoudenmire, Chris Turner, Guifre Vidal, Norman Yao, Marko Znidaric, and in particular Maksym Serbyn, for many interesting discussions and joint work on this subject.

\bibliography{mbl_annalen}
\bibliographystyle{apsrev4-1}

\end{document}